\documentclass[preprint,floatfix,showpacs,aps,nofootinbib]{revtex4}
\usepackage{graphicx}
\usepackage{dcolumn}
\usepackage{bm}
\usepackage{amsmath,amssymb}

\newcommand {\be}  {\begin{equation}}   
\newcommand {\ee}  {\end{equation}}
\newcommand {\bee} {\begin{equation*}}   
\newcommand {\eee} {\end{equation*}}
\newcommand {\bea} {\begin{eqnarray}}   
\newcommand {\eea} {\end{eqnarray}}
\newcommand {\beaa}{\begin{eqnarray*}}   
\newcommand {\eeaa}{\end{eqnarray*}}
\newcommand {\bse} {\begin{subequations}}
\newcommand {\ese} {\end{subequations}}
\newcommand {\nn}  {\nonumber}
\newcommand {\del} {\delta}

\newcommand {\pr}  {\prime}
\newcommand {\intp}{\int\frac{d^3p}{(2\pi)^3}}
\newcommand {\intpp}  {\int\frac{d^3p^\pr}{(2\pi)^3}}
\newcommand {\intppp} {\int\frac{d^3p^{\pr\pr}}{(2\pi)^3}}
\newcommand {\intx}   {\int d^3x}
\newcommand {\intxp}  {\int d^3x^\pr}
\newcommand {\onehalf}{\frac{1}{2}}
\newcommand {\vect}[1]{\mathbf{#1}}
\newcommand {\bx}   {\vect{x}}
\newcommand {\bp}   {\vect{p}}
\newcommand {\bpp}  {\bp^\pr}
\newcommand {\bppp}  {\bp^{\pr\pr}}
\newcommand {\bxp}  {\bx^\pr}
\newcommand {\simop} {U(s,s_0)}

\begin{document}
\title{Similarity flow of a neutral scalar coupled to a fixed source}

\author{Billy D. Jones} \email{bjones@apl.uw.edu}
\affiliation{Applied Physics Laboratory, University of Washington, Seattle, WA 98105}
\author{Robert J. Perry} \email{perry.6@osu.edu}  
\affiliation{Department of Physics, The Ohio State University, Columbus, OH 43210}

\begin{abstract} 

A neutral scalar meson interacting with a fixed isoscalar nucleon is evolved 
according to the similarity renormalization group. The fixed source ends up being 
too singular and an appropriately dressed static source arises from the similarity flow. 
The fixed source is dressed by a virtual cloud of mesons extending out past the inverse meson mass 
according to the Yukawa potential. For low energies the source produces a Yukawa mean field which decouples from 
the remaining meson field fluctuations. The renormalization group procedure for effective particles 
is also illustrated for this fixed-source model.

\end{abstract}
\pacs{11.10.Gh, 21.60.-n}
\maketitle

\section{Introduction}

We study fixed-source neutral scalar theory for a pedagogical introduction 
to the similarity renormalization group (SRG) \cite{G-W,SRG} and also as a 
first step to understanding nuclear structure with a fixed nucleon being dressed 
by the Yukawa mean field of the meson in ``similarity style''. The beauty of this approach   
is that the results follow as in the classic texts \cite{H-T,Wentzel}, but {\it the 
necessary regulating static source does not have to be put in by hand}; rather it follows 
from the similarity flow itself. Wegner's canonical generator \cite{Wegner, Kehrein} of  
the similarity flow is used throughout for its simplicity and 
particular interest in nuclear theory \cite{OSU08,GP08}. 

There are three approaches to the SRG: evolving the renormalization scale of the system 
according to (1) the flow equations, (2) the similarity operator, 
or (3) the renormalization group procedure for effective particles (RGPEP). 
Usually only the first or third approach is used. The utility 
of the simple model of this paper is that all three approaches can be analytically ``integrated'',
including the mostly unused second approach, and in this way the SRG method is elucidated. 
The next three sections illustrate these different approaches and 
explicitly show that they are equivalent (as is easy to formally show). The cancellations that occur 
in order to show this equivalence include precise alignment between 
the free quadratic, interacting linear source, and singular second order background energy terms.

\section{Flow equations}
\label{sec:flow} 

The canonical Hamiltonian for a relativistic neutral scalar of mass $m$ 
interacting with a static source $\rho(\bx)$ with coupling constant $g$ is 
\be
H_{can} = \onehalf \intx \left[
\pi^2(\bx) + \left|\nabla\phi(\bx)\right|^2 + 
m^2\phi^2(\bx) + 2\,g \,\rho(\bx)\phi(\bx)  
\right] \label{eq:hcan}
\ee
As is well known \cite{H-T,Wentzel}, a fixed source [$\rho(\bx)=\del^3(x)$] is too singular 
for this problem and leads to a linear divergence in the energy of the system. 
Thus the fixed source has to be dressed and in general becomes a static source with 
specific form, as will be shown, following from the similarity flow. 

To set up the flow equations, it is convenient to use momentum space: 
\bse
\label{eq:ms}
\bea
\phi(\bx)&=&\intp e^{i\bp\cdot\bx}\phi(\bp)\label{eq:2a}\\
\pi(\bx)&=&\intp e^{-i\bp\cdot\bx}\pi(\bp)\label{eq:2b}\\
\rho(\bx)&=&\intp e^{i\bp\cdot\bx}\rho(\bp)\label{eq:2c}
\eea
\ese
where $\phi$ and $\pi$ are quantum field operators ($\phi$ is the scalar field and $\pi$ its 
conjugate momentum field) and $\rho$ is  
a classical field. In general, $\phi(\bx)$, $\pi(\bx)$, and $\rho(\bx)$ are real in this model 
but not so in momentum space.\footnote{It will turn out that $\rho(\bp)$ is a real even function of $\bp$ in this model. 
But initially we are keeping the expressions general so as to not have circular logic.}  
Note the opposite phase convention between the definitions of $\phi(\bp)$ and $\pi(\bp)$; 
this is standard and convenient for the momentum space canonial commutation relations that follow as shown below. 
Plugging these definitions into Eq.~(\ref{eq:hcan}), the canonical Hamiltonian 
becomes
\be
H_{can} = \onehalf \intp \left[
\pi^\dag(\bp)\pi(\bp) + E_\bp^2 \phi^\dag(\bp)\phi(\bp) + g \,\rho(\bp)\phi^\dag(\bp) + g \,\rho^*(\bp)\phi(\bp)
\right] \label{eq:hcanp}
\ee
where $E_\bp = \sqrt{\bp^2+m^2}$ and since $\phi(\bx)$, $\pi(\bx)$, and $\rho(\bx)$ are real, 
we have used $\phi(-\bp)=\phi^\dag(\bp)$, $\pi(-\bp)=\pi^\dag(\bp)$, and $\rho(-\bp)=\rho^*(\bp)$ respectively. 
In this paper we use fields $\phi$ and $\pi$ in momentum or position space for direct calculations, however 
in terms of Fock-space creation and annihilation operators, $a^\dag(\bp)$ and $a(\bp)$ respectively, 
the scalar field and its conjugate momentum field are given by 
Eqs.~(\ref{eq:2a}) and (\ref{eq:2b}) respectively with
\bse
\bea
\phi(\bp) = \frac{a(\bp) + a^\dag(-\bp)}{2 E_\bp}\label{eq:particlecontent}\\
\pi(\bp) = \frac{a(-\bp) - a^\dag(\bp)}{2 i}
\eea
\ese
using a relativistic norm. 
Using these expressions, the particle content of any of our results that follow can be readily obtained. 

The canonical commutation relations are given by
\bse
\label{eq:ccr}
\bea
\left[\phi(\bx),\pi(\bx^\pr)\right]&=&i\,\del^3(x-x^\pr)\\
\left[\phi(\bx),\phi(\bx^\pr)\right]&=&\left[\pi(\bx),\pi(\bx^\pr)\right]=0
\eea
\ese
which in momentum space become
\bse
\label{eq:ccrp}
\bea
\left[\phi(\bp),\pi(\bp^\pr)\right]&=&i\,(2\pi)^3\,\del^3(p-p^\pr)\label{eq:ccrpA}\\
\left[\phi(\bp),\phi(\bp^\pr)\right]&=&\left[\pi(\bp),\pi(\bp^\pr)\right]=0
\eea
\ese
To handle products of fields, note that for 
commutators of general Fock-space operator objects $A$, $B$, and $C$, the following useful relations hold  
\bse
\label{eq:usefulABC}
\bea
{[}AB,C]&=&[A,C]B+A[B,C]\label{eq:ableft}\\
{[}A,BC]&=&[A,B]C+B[A,C]\label{eq:bcright}
\eea
\ese
with the uncontracted objects both on the inside by the plus sign, or with the opposite ordering of the two terms, both on the outside 
away from the commutators (there is no mixed ``inside-outside'' case). 

The {\it similarity flow equation} \cite{Walhout} for effective Hamiltonian $H_s$ at scale $s$,
\be
\frac{dH_s}{ds}=\left[\eta_s,H_s\right]\label{eq:simfloweq}
\ee
using Wegner's {\it canonical generator} \cite{Kehrein} (with $H_s\equiv H_0+V_s$),
\be
\label{eq:eta}
\eta_s\equiv \left[H_0,V_s\right]
\ee
is given by
\be
\frac{dH_s}{ds}=\left[\left[H_0,V_s\right],H_0\right]+\left[\left[H_0,V_s\right],V_s\right]
\label{eq:floweq}
\ee
containing a linear and quadratic term in $V_s$ (that is it to all orders).  
The effective Hamiltonian has been divided according to
\be
H_s\equiv H_0+V_s
\ee
with $H_0$ the free Hamiltonian and $V_s$ the effective interaction, however this division 
can be made quite arbitrarily in general (always maintaining ``energy band-diagonalness'') 
and if there is a best division or not 
is an active open question \cite{OSU08,GP08}. For the model of this paper, the free 
and energy-diagonal (including interactions that are diagonal in energy space) Hamiltonians are one 
and the same so the issue does not come up further. 

As easily seen from Eq.~(\ref{eq:floweq}), the dimensions of $s$ are inverse energy squared or position squared, 
an effective size \cite{RGPEP11} at which the system is probed. 
In general the domain of $s$ is $0 \leq s \leq \infty$ with $s\rightarrow 0$ the high-energy limit 
and $s\rightarrow\infty$ the low-energy limit. 
For fixed-source neutral scalar theory we define  
\bse
\label{eq:theham}
\bea
H_0 &=& \onehalf \intp \left[ \pi^\dag(\bp)\pi(\bp) + E_\bp^2 \phi^\dag(\bp)\phi(\bp) \right] \label{eq:H0}\\
V_s &=& \frac{\Sigma_s}{\sqrt{s}}+\frac{g}{2} \intp 
\left[\rho_s(\bp)\phi^\dag(\bp)+\rho_s^*(\bp)\phi(\bp)\right] \label{eq:Vs}
\eea
\ese
where the static source $\rho_s(\bp)$ has been allowed to run with $s$, and an overall dimensionless background energy $\Sigma_s$ has 
been added as will be shown to be required in what follows in order for the similarity flow equation to close on itself. 
It will also be shown that the static source 
$\rho_s(\bx)$ in position space flows to a delta function in the high-energy limit $s\rightarrow 0$, so indeed 
we are studying fixed-source neutral scalar theory and it gets dressed as the system flows towards lower energies (larger $s$). 

We now show that Eq.~(\ref{eq:theham}) is closed under the similarity flow equation, and in the process 
obtain the flow equations for $\rho_s(\bp)$ and $\Sigma_s$. Then these flow equations are integrated and their 
representation in position space is discussed for elucidation. 

Similarity flow calculations start with a derivation of the generator $\eta_s$. 
In this model, given Eqs.~(\ref{eq:eta}) and (\ref{eq:theham}), $\eta_s$ 
becomes a big commutator: 
\bea
\eta_s&=&\left[H_0,V_s\right]\nn\\
&=& \frac{g}{4}\intp \intpp \left[ 
\pi^\dag(\bp)\pi(\bp) + E_\bp^2 \phi^\dag(\bp)\phi(\bp),
\rho_s(\bpp)\phi^\dag(\bpp)+\rho_s^*(\bpp)\phi(\bpp)\right]
\label{eq:etas1}
\eea
where we have noted that $\Sigma_s$ is a c-number and commutes with everything. Using the above 
useful relations and canonical commutation relations it is easy to show that Eq.~(\ref{eq:etas1}) 
becomes 
\be
\eta_s = \frac{-ig}{2}\intp\left[ \rho_s(\bp)\pi(\bp) + \rho_s^*(\bp)\pi^\dag(\bp) \right]
\label{eq:etas2}
\ee

The next step is to insert this result for $\eta_s$ into the two terms of the right-hand side of 
Eq.~(\ref{eq:floweq}). The first term becomes
\bee
\left[\eta_s,H_0\right]=\frac{-ig}{4}\intp \intpp \left[ 
\rho_s(\bp)\pi(\bp) + \rho_s^*(\bp)\pi^\dag(\bp),
\pi^\dag(\bpp)\pi(\bpp) + E_{\bpp}^2 \phi^\dag(\bpp)\phi(\bpp)\right]
\eee
which similar to the $\eta_s$ derivation after some simple algebra becomes 
\be
\left[\eta_s,H_0\right] = \frac{-g}{2}\intp E_\bp^2\left[ \rho_s(\bp)\phi^\dag(\bp) + \rho_s^*(\bp)\phi(\bp) \right]
\label{eq:etaH0term}
\ee
Note that except for a minus sign and the factor of $E_\bp^2$ 
this is the same integral as the one in the effective interaction of Eq.~(\ref{eq:Vs}); thus this gives the 
flow equation for the static source. Before writing this, first we need to discuss 
the second term of Eq.~(\ref{eq:floweq}):
\bee
\left[\eta_s,V_s\right]=\frac{-ig^2}{4}\intp \intpp \left[ 
\rho_s(\bp)\pi(\bp) + \rho_s^*(\bp)\pi^\dag(\bp),
\rho_s(\bpp)\phi^\dag(\bpp) + \rho_s^*(\bpp)\phi(\bpp)\right]
\eee
where once again $\Sigma_s$ canceled in the commutator since it is not an operator. After using the 
canonical commutation relations and $\rho_s(-\bp)=\rho_s^*(\bp)$  this reduces to
\be
\left[\eta_s,V_s\right] = -g^2\intp \rho^*_s(\bp) \rho_s(\bp)
\label{eq:etaVsterm}
\ee
which will be shown to be the well-known background energy shift of this model. 
Now we use these $\eta_s$ commutator results to derive the flow equations. 

These just calculated $\eta_s$ commutators are the right-hand side of Eq.~(\ref{eq:floweq}); now we work on its 
left-hand side. Thus we explicitly take a derivative of Eq.~(\ref{eq:theham}) with respect to scale $s$ 
and set it equal to the sum of Eqs.~(\ref{eq:etaH0term}) and (\ref{eq:etaVsterm}). 
This shows that the effective Hamiltonian of Eq.~(\ref{eq:theham}) is indeed a closed structure 
and that its flow equations are 
\bse
\bea
\frac{d\rho_s(\bp)}{ds}&=&-E_\bp^2\, \rho_s(\bp)\\
\frac{d}{ds}\left(\frac{\Sigma_s}{\sqrt{s}}\right)&=&-g^2\intp \rho^*_s(\bp) \rho_s(\bp)
\label{eq:sigmaflow}
\eea
\ese
There are the only flow equations of this model. There are two, therefore two initial conditions are required to define the model. 
Integrating this first flow equation gives 
\be
\rho_s(\bp)= \rho_{s_0}(\bp) \,e^{-(s-s_0)E_\bp^2}
\label{eq:rhos1}
\ee
where recall $E_\bp=\sqrt{\bp^2+m^2}$ and we have $s \geq s_0 \geq 0$. 
In order to obtain the explicit momentum dependence of this dressed source we 
postulate that at high energies, $s_0\rightarrow 0$, a finite momentum $\bp$ is 
completely negligible (compared to $1/\sqrt{s_0}$) thus giving a momentum independent limit:
\be
\lim_{s_0\rightarrow 0} \rho_{s_0}(\bp) \equiv 1
\ee
with unity chosen so as to match onto the delta function starting point in position space. 
This is the first initial condition required to define the model. It gives 
a delta function (fixed) source in postion space at high energies. 
Plugging this initial condition into Eq.~(\ref{eq:rhos1}) gives the most useful  
form of our dressed source with all momentum dependence explicitly shown:
\be
\rho_s(\bp)= e^{-sE_\bp^2}
\label{eq:rhos2}
\ee
We stress that this functional form was not chosen but is a solution of the similarity flow equation. 
As already mentioned in a footnote, note that this form in momentum space is 
even in $\bp$, $\rho_s(-\bp)=\rho_s(\bp)$, and therefore when combined with $\rho_s(-\bp)=\rho_s^*(\bp)$ from the reality 
of $\rho_s$ in position space, we have that $\rho_s$ must be real in momentum space as well: $\rho_s(\bp)=\rho_s^*(\bp)$. 
This will be used here on out to simplify expressions. 
Before looking at these results in position space, 
first we integrate the second flow equation for $\Sigma_s$ which uses 
this result from the first flow equation for $\rho_s(\bp)$.

Inserting Eq.~(\ref{eq:rhos2}) into Eq.~(\ref{eq:sigmaflow}) and then integrating over momentum and scale $s$ gives
\be
\Sigma_s=I_1 \sqrt{m^2 s}+\frac{g^2}{8\pi}  
\left[ \frac{e^{-2m^2s}}{\sqrt{2\pi}} - \sqrt{m^2 s}\,{\rm erfc}\left(\sqrt{2m^2s}\right) \right]
\label{eq:Sigmas}
\ee
where $I_1$ is a dimensionless constant of integration 
and ${\rm erfc}$ is the well-known complementary error function. 
$I_1$ is the second required initial condition of the model. Often 
\cite{H-T,Wentzel} $I_1\equiv 0$ for aesthetic reasons; 
this is the choice that in the low-energy limit, $s\rightarrow\infty$, $\Sigma_s$ vanishes exponentially and thus 
does not appear explicitly in the effective low-energy theory. 
No (non gravitational) physical results depend on this choice for $I_1$ because $\Sigma_s/\sqrt{s}$ is just an overall 
scale-dependent but otherwise constant shift of the energy scale of the system. 
Eq.~(\ref{eq:Sigmas}) is the well-known linear divergence of the background energy in this model although it might not 
be clear that that is what we have. To gain insight we thus look at things in position space. 
First the static source itself and then this background energy. 

In position space the meson cloud that dresses the fixed nucleon is given by the Fourier transform 
of the derived dressed source, Eq.~(\ref{eq:rhos2}), which is a Gaussian in momentum space and thus a Gaussian 
in position space as well:   
\bea
\rho_s(\bx)&=&\intp e^{i\bp\cdot\bx}\rho_s(\bp)\nn\\
&=&\intp e^{i\bp\cdot\bx}e^{-sE_\bp^2}\nn\\
&=&\frac{e^{-m^2s}}{2\pi^2 r} \int_0^\infty dp \, p\sin(pr)e^{-s(p^2+m^2)}\nn\\
&=&\frac{e^{-m^2s}e^{-\frac{r^2}{4s}}}{8(\pi s)^{\frac{3}{2}}}
\label{eq:rhosx}
\eea
where $r = |\bx|$. Indeed the high-energy limit gives us back the fixed source we started with:
\be
\lim_{s\rightarrow 0} \rho_s(\bx)=\del^3(x)
\ee
The similarity flow has produced a dressed fixed source for this model. 
For high energies, $m^2s< 1$, it flows to a static source of size 
$r\sim\sqrt{s}$. This is a nice result: the meson cloud extends out to $1/m$ exponentially damped thereafter, and for 
$\sqrt{s}$ inside of this then it is the effective size of the system that is being resolved. 
For low energies, $m^2s> 1$, the source exponentially decouples from the scalar meson and in the 
extreme low energy limit, the similarity flow diagonalizes the interaction away:
\be
\lim_{s\rightarrow \infty} \rho_s(\bx)=0
\ee
 
Now we discuss the background energy $\Sigma_s$ in position space. 
$\Sigma_s$ is a functional of the dressed source just discussed. This 
follows from the $\rho_s(\bp)$ on the right-hand side of the $\Sigma_s$ flow equation, Eq.~(\ref{eq:sigmaflow}).  
To see the picture in position space, integrate Eq.~(\ref{eq:sigmaflow}), but this time integrate 
over just scale $s$ using the dressed source result of Eq.~(\ref{eq:rhos2}). 
Integrating Eq.~(\ref{eq:sigmaflow}) from an arbitrary initial scale $s_0$ 
to some final scale $s$ with $s\geq s_0\geq 0$ gives 
\bea
\frac{\Sigma_s}{\sqrt{s}}&=&\frac{\Sigma_{s_0}}{\sqrt{s_0}}-g^2\int_{s_0}^sds^\pr \intp \rho^*_s(\bp) \rho_s(\bp)\nn\\
&=&\frac{\Sigma_{s_0}}{\sqrt{s_0}}-g^2\int_{s_0}^sds^\pr \intp e^{-2sE_\bp^2}\nn\\
&=&\frac{\Sigma_{s_0}}{\sqrt{s_0}}-g^2\intp \frac{1}{2E_\bp^2}\left[\rho^*_{s_0}(\bp) \rho_{s_0}(\bp)-\rho^*_s(\bp) \rho_s(\bp)\right]
\label{eq:sigmasSfirst1}
\eea
where for aesthetic reasons we have inserted Eq.~(\ref{eq:rhos2}) back into the result after performing the explicit integral over $s$. 
Note that the integral over $s$ has produced a factor of $1/E_\bp^2=1/(\bp^2+m^2)$. This is 
where the Yukawa potential that arises comes from due to the well-known Fourier transform 
of this factor:
\be
\intp \frac{1}{\bp^2+m^2}e^{i\bp\cdot(\bx-\bxp)}=\frac{e^{-m|\bx-\bxp|}}{4\pi|\bx-\bxp|}
\label{eq:yukawaIntegral}
\ee
Continuing, using the inverse relation of Eq.~(\ref{eq:2c}) inserted into Eq.~(\ref{eq:sigmasSfirst1}) 
gives
\be
\frac{\Sigma_s}{\sqrt{s}}=\frac{\Sigma_{s_0}}{\sqrt{s_0}}-g^2\intp 
\frac{1}{2E_\bp^2}\intx\intxp e^{i\bp\cdot(\bx-\bxp)}\left[\rho_{s_0}(\bx)\rho_{s_0}(\bxp)-\rho_s(\bx)\rho_s(\bxp)\right]
\ee
Finally, integrating over momentum, we recognize exactly the Yukawa potential integral of Eq.~(\ref{eq:yukawaIntegral}). 
Thus, altogether we have  
\be
\frac{\Sigma_s}{\sqrt{s}}=\frac{\Sigma_{s_0}}{\sqrt{s_0}}-\frac{g^2}{2} 
\intx\intxp 
\frac{e^{-m|\bx-\bxp|}}{4\pi|\bx-\bxp|}
\left[\rho_{s_0}(\bx)\rho_{s_0}(\bxp)-\rho_s(\bx)\rho_s(\bxp)\right]
\label{eq:posspace1}
\ee
with source $\rho_s(\bx)$ given explicitly by Eq.~(\ref{eq:rhosx}).
Recall $s\geq s_0\geq 0$, with $s\rightarrow s_0$ the high-energy limit and $s\rightarrow\infty$ the low-energy limit.
We see that $\Sigma_s$ is a background energy term with 
sources interacting via the Yukawa potential out to a range of order the Compton wavelength of the meson 
and then exponentially damped beyond that. In addition, the 
effective range of interaction of the sources, for a particular $\Sigma_s$ at scale $s$,
is also restricted to be between $r\sim\sqrt{s_0}$ out to $r\sim\sqrt{s}$. The scale $s$ 
(strictly speaking $\sqrt{s}$ with our current conventions) sets the size at which the system is probed. 

\section{Similarity operator}
\label{sec:op}

The previous section is a follow-up to the presentation in the static source chapter of 
the classic text of Henley and Thirring \cite{H-T}. We showed the similarity flow of the model 
leads to a particular static source with background energy Yukawa 
interactions that shift the overall energy scale of the system in a well-prescribed way. The static source 
did not have to be put in by hand, but rather followed from the similarity flow equation. 
The present section is a follow-up to Wentzel's unitary transformation 
discussed at the end of the ``Real Field with Sources'' section of \cite{Wentzel}.  
In the current and next sections, we 
use fixed-source neutral scalar theory to study approaches of the SRG that have not been integrated before 
in closed form, thus aiding understanding and helping to establish more confidence in all of the approaches: 
flow equations, similarity operator, and RGPEP.   

Since the start of the SRG \cite{SRGstart}, two Hamiltonians at different scales have been related by 
a similarity operator $U(s,s_0)$ as in (using current notations)
\be
H_s = U(s,s_0)H_{s_0}\,U^\dag(s,s_0)
\label{eq:simopeq}
\ee
where $U(s,s_0)$ is unitary:
\be
U(s,s_0)U^\dag(s,s_0)=U^\dag(s,s_0)U(s,s_0)=1
\label{eq:unitary}
\ee
with $s\geq s_0\geq 0$. 
But this similarity operator is usually not used in direct calculations, but rather  
flow equations are established as in the previous section which are more amenable to numerical investigations. 
In this section, we will directly calculate $\simop$ for fixed-source neutral scalar theory and show that 
it is a regulated version of the classic one Wentzel uses in \cite{Wentzel} to diagonalize the same Hamiltonian. Then we will use this derived 
$\simop$ to explicitly evolve the Hamiltonian from scale $s_0$ to $s$ and show that 
the same $H_s$ arises as in the previous section. Then we will use this same $\simop$ to evolve the scalar field 
and discuss how a Yukawa mean field arises at low energies. In the process of the calculations, 
much is learned through the exact cancellations between 
precisely aligned free (that become interacting to cancel), interacting, and singular background energy terms. 

It is easy to formally establish the equivalence between the flow equations approach, Eq.~(\ref{eq:floweq}), and 
the similarity operator approach, Eq.~(\ref{eq:simopeq}). For completeness we present this formal argument, 
and then we analytically verify the equivalence in the bulk of this section. The similarity operator 
approach starts with a definition of $\simop$ in terms of 
the similarity generator $\eta_s$ of the previous section. We show that the two approaches are equivalent if the 
similarity operator $\simop$ is given by the following Dyson series 
\be
U(s,s_0) = {\cal S}\,exp\left[\int_{s_0}^s \eta_{s^\pr}ds^\prime\right]
\label{eq:simop}
\ee
where ${\cal S}$ is the 
scale-ordering operator that orders operators from {\it right} to {\it left} in order 
of increasing scale $s$ (a strict analogy with the usual time-ordering operator), 
and $\eta_s$ is the same similarity generator as derived in the previous section: 
defined by Eq.~(\ref{eq:eta}) with result Eq.~(\ref{eq:etas2}). We are being careful with ordering 
because $\eta_s$ is a Fock-space operator which does not necessarily commute for different 
values of $s$.  ${\cal S}$ is not required for the model of this paper, but we nevertheless mention it 
so that this similarity operator defining equation works for the general case.  

To show the formal equivalence between the flow equations and similarity operator approaches, 
we need to take a derivative of Eq.~(\ref{eq:simopeq}) with respect to $s$ and show that it leads 
to the same right-hand side as Eq.~(\ref{eq:simfloweq}). 
First, since the Dyson series of Eq.~(\ref{eq:simop}) is easy to differentiate we have two general  
results that will be used in what follows: 
\bse
\bea
\frac{dU(s,s_0)}{ds} &=& \eta_s\,U(s,s_0)\label{eq:d1}\\
\frac{dU^\dagger(s,s_0)}{ds} &=& -U^\dagger(s,s_0)\,\eta_s\label{eq:d2}
\eea
\ese
This second result uses the facts that $\eta_s$ is antihermitian, $\eta_s^\dag=-\eta_s$, with our current conventions 
and that ${\cal S}^\dagger$ [the scale-ordering operator for $U^\dag(s,s_0)$] 
orders operators from {\it left} to {\it right}, 
oppositely to that of ${\cal S}$, which follows easily from the usual properties of hermitian conjugation.
Given these two results it is easy to take a derivative of Eq.~(\ref{eq:simopeq}) 
using the product rule of differentiation: 
\bea
\frac{dH_s}{ds} &=& \frac{dU(s,s_0)}{ds}H_{s_0}U^\dagger(s,s_0)+
U(s,s_0)H_{s_0}\frac{dU^\dagger(s,s_0)}{ds}\nn\\
&=&\eta_s\,U(s,s_0)H_{s_0}U^\dagger(s,s_0)-U(s,s_0)H_{s_0}U^\dagger(s,s_0)\,\eta_s\nn\\
&=&[\eta_s,H_s]
\label{eq:minussign}\eea
and we indeed end up with Eq.~(\ref{eq:simfloweq}) as was to be shown. Note that this formal equivalence 
did not require a specific form for $\eta_s$ although our explicit functional operator calculation below 
does use Eq.~(\ref{eq:etas2}), the similarity generator for the model of this paper.

Before leaving this formal discussion we show that the spectrum of $H_s$ is independent of $s$ while its 
eigenstates are not. This is easy to show since $\simop$ is unitary. Multiplying the eigenvalue equation for 
$H_{s_0}$ by $U(s,s_0)$ and using unitarity gives
\bea
H_{s_0} |\Psi_{s_0}\rangle&=& E |\Psi_{s_0}\rangle\\
\Longrightarrow~~~~~
\underbrace{U(s ,s_0) H_{s_0} U^\dagger(s ,s_0)}_{H_s}
\underbrace{ U(s, s_0)|\Psi_{s_0}\rangle}_{|\Psi_s\rangle}&=&E \underbrace{U(s, s_0) 
|\Psi_{s_0}\rangle}_{|\Psi_s\rangle}
\label{eq:schro}
\eea
showing identical scale-independent spectra for $H_s$ and $H_{s_0}$ as long as their eigenstates do depend on $s$ by a  
multiplication with the similarity operator. 
This unitary transformation keeps the full state space of the initial Hamiltonian $H_{s_0}$. No states are removed, they are just 
resolved at different size scales ($\sqrt{s_0}$ and the larger $\sqrt{s}$ respectively) 
and the Hamiltonian is actively ``rotated'' from scale $s_0$ to $s$.

Now we explicitly show that evolution with the similarity operator, 
the right-hand side of Eq.~(\ref{eq:simopeq}), leads to the same effective Hamiltonian 
$H_s$ as derived in the previous section from the flow equations. 
First, we write the expression for the similarity operator itself and then use it to evolve 
the initial Hamilotonian $H_{s_0}$.  The similarity operator $U(s,s_0)$ is given by Eq.~(\ref{eq:simop}) 
with $\eta_s$ replaced by Eq.~(\ref{eq:etas2}):    
\bea
U(s,s_0) &=&  {\cal S}\,exp\left[\frac{-ig}{2}\int_{s_0}^s ds^\pr\intp\left\{ \rho_{s^\pr}(\bp)\pi(\bp) + 
\rho_{s^\pr}^*(\bp)\pi^\dag(\bp) \right\}\right]\nn\\
&=&exp\left[-ig\int_{s_0}^s ds^\pr\intp \rho_{s^\pr}(\bp)\pi(\bp)\right]
\label{eq:uintermsofpi}
\eea 
where as already mentioned, since $\eta_s$ commutes with itself at different scales $s$ 
(because the conjugate momentum field $\pi(\bp)$ commutes with itself and its dagger), 
the scale-ordering operator ${\cal S}$ is not required and is thus dropped from further discussions in this paper.  
Also, the two terms of $\eta_s$ were combined since $\rho_s(\bp)$ is real and even 
and $\pi^\dag(\bp)=\pi(-\bp)$ as previously discussed.   
Continuing, the initial Hamiltonian $H_{s_0}$ 
is given by the sum of Eqs.~(\ref{eq:H0}) and (\ref{eq:Vs}) with $s$ set to $s_0$. Thus, 
the right-hand side of Eq.~(\ref{eq:simopeq}) in all its gory detail becomes
\bea
&&\!\!\!\!\!\!\!\!\!\!\!\!\!\!\!\!\!\!\!\!\!\!\!\!\!\!\!\!\!U(s,s_0)H_{s_0}U^\dag(s,s_0)\nn\\
&=& exp\left[-ig\int_{s_0}^s ds^\pr\intpp \rho_{s^\pr}(\bpp)\pi(\bpp)\right]\nn\\
&\times& \left[\frac{\Sigma_{s_0}}{\sqrt{s_0}}+\onehalf \intp \left\{ \pi^\dag(\bp)\pi(\bp) + E_\bp^2 \phi^\dag(\bp)\phi(\bp) \right\} 
+g\intp \rho_{s_0}(\bp)\phi(\bp)\right]\nn\\
&\times& exp\left[ig\int_{s_0}^s ds^{\pr\pr}\intppp \rho_{s^{\pr\pr}}(\bppp)\pi(\bppp)\right]
\label{eq:simoprhsStart}
\eea
where once again $\rho_s(\bp)$ being real and even along with $\phi^\dag(\bp)=\phi(-\bp)$ was used to combine the two interaction terms of 
Eq.~(\ref{eq:Vs}). 

In order to manipulate Eq.~(\ref{eq:simoprhsStart}) further, first we need to discuss a well-known result from quantum mechanics, 
appropriately generalized for functionals over momentum such as the similarity operator for this model: $U[\pi(\bp)]$. 
Given the canonical commutation relation of Eq.~(\ref{eq:ccrpA}) we have
\bea
\left[\phi(\bp),U^\dag[\pi(\bpp)]\right]&=&\intppp[\phi(\bp),\pi(\bppp)]\frac{\del}{\del\pi(\bppp)}U^\dag[\pi(\bpp)]\nn\\
&=& i \frac{\del}{\del\pi(\bp)}U^\dag[\pi(\bpp)]\nn\\
&=&-g\,U^\dag(s,s_0)\int_{s_0}^s ds^\pr \rho_{s^\pr}(\bp)
\label{eq:funcdiff}
\eea
where the `$\del/\del\pi(\bp)$' signifies functional differentiation and the last step required functionally differentiating 
the dagger of Eq.~(\ref{eq:uintermsofpi}) and respectively multiplying by $i$. 
We mention one further simplification used in what follows to ``move past $U^\dag$''. The general terms that appear can be written as 
\bea
U\times\left({\cal O}U^\dag\right)&=&U\times\left(U^\dag {\cal O}+[{\cal O},U^\dag]\right)\nn\\
&=&{\cal O}\,+\,U\!\times\![{\cal O},U^\dag]
\label{eq:abba}
\eea
where ${\cal O}$ is an arbitrary product of quantum fields and unitarity of $U$ is used in the last step. A product 
in ${\cal O}$ is reduced further in that final commutator $[{\cal O},U^\dag]$ by using the useful relations of Eq.~(\ref{eq:usefulABC}) 
and this leads to the lowest level result Eq.~(\ref{eq:funcdiff}) as already discussed.  Given these results, 
the evaluation of Eq.~(\ref{eq:simoprhsStart}) is straightforward which we now show.

There are four terms in Eq.~(\ref{eq:simoprhsStart}) which for short we call the 
$\Sigma_{s_0}$, $\pi^\dag\pi$, $\phi^\dag\phi$, and $g\phi$ terms. We will handle each in turn. The $\Sigma_{s_0}$ and 
$\pi^\dag\pi$ terms trivially commute with $U^\dag$ because $\Sigma_{s_0}$ is a c-number and any functional of $\pi(\bp)$ commutes with $U^\dag[\pi(\bp)]$. 
This leaves the $\phi^\dag\phi$ and $g\phi$ terms. Given the algebra discussion of the previous paragraph, 
along with the symmetry discussions of $\rho_{s}(\bp)$ and $\phi(\bp)$, the $\phi^\dag\phi$ and $g\phi$ terms follow simply and both 
contain interacting pieces: 
\bse
\label{eq:twoterms}
\bea 
\label{eq:twotermsA}
U(s,s_0)\onehalf \intp E_\bp^2 \phi^\dag(\bp)\phi(\bp) U^\dag(s,s_0)&=& \onehalf \intp E_\bp^2 \phi^\dag(\bp)\phi(\bp)\\
&-&\frac{g}{2}\intp E_\bp^2\left[\phi(\bp)+\phi_s(\bp)\right]\int_{s_0}^s ds^\pr \rho_{s^\pr}(\bp)\nn\\
U(s,s_0)g\intp \rho_{s_0}(\bp)\phi(\bp) U^\dag(s,s_0)&=&g\intp \rho_{s_0}(\bp)\phi_s(\bp)
\label{eq:twotermsB}
\eea
\ese
where the running similarity field, $\phi_s(\bp)$, is given by  
\bea
\phi_s(\bp)\equiv
U(s,s_0)\phi(\bp)U^\dag(s,s_0)&=& U(s,s_0)\left\{U^\dag(s,s_0)\phi(\bp)+\left[\phi(\bp),U^\dag(s,s_0)\right]\right\}\nn\\
&=&\phi(\bp)-g\int_{s_0}^s ds^\pr \rho_{s^\pr}(\bp)
\label{eq:42}
\eea
in terms of an arbitrary static source $\rho_{s}(\bp)$  even in $\bp$. 
Substituting Eq.~(\ref{eq:42}) into Eqs.~(\ref{eq:twotermsA}) and (\ref{eq:twotermsB}), 
combining with the $\Sigma_{s_0}$ and $\pi^\dag\pi$ terms, Eq.~(\ref{eq:simoprhsStart}) altogether becomes 
\bea
U(s,s_0)H_{s_0}U^\dag(s,s_0)&=& \frac{\Sigma_{s_0}}{\sqrt{s_0}}+\onehalf \intp 
\left[ \pi^\dag(\bp)\pi(\bp) + E_\bp^2 \phi^\dag(\bp)\phi(\bp) \right]\nn\\
&-&\frac{g}{2}\intp E_\bp^2\left[2\,\phi(\bp)-g\int_{s_0}^s ds^\pr \rho_{s^\pr}(\bp)\right]\int_{s_0}^s ds^{\pr\pr} \rho_{s^{\pr\pr}}(\bp)\nn\\
&+&g\intp \rho_{s_0}(\bp)\left[\phi(\bp)-g\int_{s_0}^s ds^\pr \rho_{s^\pr}(\bp)\right]
\label{eq:almostA}
\eea
To proceed, Eq.~(\ref{eq:rhos2}) for this fixed-source similarity model implies
\be
\int_{s_0}^s ds^\pr \rho_{s^\pr}(\bp)=\frac{1}{E_\bp^2}\left[\rho_{s_0}(\bp)-\rho_{s}(\bp)\right]
\label{eq:44}
\ee
and thus Eq.~(\ref{eq:almostA}) becomes
\bea
U(s,s_0)H_{s_0}U^\dag(s,s_0)&=& \frac{\Sigma_{s_0}}{\sqrt{s_0}}+\onehalf \intp 
\left[ \pi^\dag(\bp)\pi(\bp) + E_\bp^2 \phi^\dag(\bp)\phi(\bp) \right]\nn\\
&-&\frac{g}{2}\intp\left\{\rho_{s_0}(\bp)-\rho_{s}(\bp)\right\}  \left[2\,\phi(\bp)-\frac{g}{E_\bp^2}\left\{\rho_{s_0}(\bp)-\rho_{s}(\bp)\right\}\right]\nn\\
&+&g\intp \rho_{s_0}(\bp)\left[\phi(\bp)-\frac{g}{E_\bp^2}\left\{\rho_{s_0}(\bp)-\rho_{s}(\bp)\right\}\right]
\label{eq:almostB}
\eea
Combining, beautiful cancellations occur between the remaining $-\frac{g}{2}\phi^\dag\phi$ and $g\phi$ terms, 
and also we see the form of $\Sigma_s$ from  Eq.~(\ref{eq:sigmasSfirst1}) appearing; all told this becomes 
\bea
U(s,s_0)H_{s_0}U^\dag(s,s_0) &=& \onehalf \intp \left[ \pi^\dag(\bp)\pi(\bp) + E_\bp^2 \phi^\dag(\bp)\phi(\bp) \right]\nn\\
&&~~~~~~~~~~+ \frac{\Sigma_s}{\sqrt{s}}+g\intp\rho_s(\bp)\phi(\bp)
\label{eq:theham2}
\eea
which is equivalent to the sum of the two pieces of Eq.~(\ref{eq:theham}) as was to be demonstrated.

In terms of the running similarity field $\phi_s(\bp)$ of Eq.~(\ref{eq:42}), from the proof that we just went through, 
with unitary $U(s,s_0)$, 
it is clear that Eq.~(\ref{eq:theham2}) is also equivalent to this form of the effective Hamiltonian
\bea
U(s,s_0)H_{s_0}U^\dag(s,s_0) &=& \onehalf \intp \left[ \pi^\dag(\bp)\pi(\bp) + E_\bp^2 \phi_s^\dag(\bp)\phi_s(\bp) \right]\nn\\
&&~~~~~~~~~~+ \frac{\Sigma_{s_0}}{\sqrt{s_0}}+g\intp\rho_{s_0}(\bp)\phi_s(\bp)
\label{eq:theham3}
\eea
with the scales $s_0$ and $s$ carefully placed in this equation. 
So one can think in terms of a dressed source and background energy ($\rho_s(\bp)$ and $\Sigma_s$ respectively) as in 
Eq.~(\ref{eq:theham2}) or in terms of a running similarity field $\phi_s(\bp)$ as in Eq.~(\ref{eq:theham3}). Either way  
leads to the same result for the effective Hamiltonian $H_s$. In addition, the following section on RGPEP is a third 
equivalent form of the effective Hamiltonian that could be useful for connecting effective low-energy 
particles with the original short-distance physics. 
We close this section by looking at the running similarity field in position space.

Recall the form of the background energy $\Sigma_s$ in position space, Eq.~(\ref{eq:posspace1}), because   
the running similarity field in position space, $\phi_s(\bx)$, follows the same physics: dressing by virtual mesons via 
the Yukawa potential. Thus, the running similarity field in position space, from Eq.~(\ref{eq:42}) with Eq.~(\ref{eq:44}) inserted, becomes
\bea
\phi_s(\bx)\equiv
U(s,s_0)\phi(\bx)U^\dag(s,s_0)&=&
\intp e^{i\bp\cdot\bx}\phi_s(\bp)\nn\\
&=&\intp e^{i\bp\cdot\bx}\left[\phi(\bp)-\frac{g}{E_\bp^2}\left\{\rho_{s_0}(\bp)-\rho_{s}(\bp)\right\}\right]\nn\\
&=&\phi(\bx)-g\intp \frac{e^{i\bp\cdot\bx}}{E_\bp^2}\intxp e^{-i\bp\cdot\bxp}\left[\rho_{s_0}(\bxp)-\rho_{s}(\bxp)\right]\nn\\
&=&\phi(\bx)-g\intxp \frac{e^{-m|\bx-\bxp|}}{4\pi|\bx-\bxp|}\left[\rho_{s_0}(\bxp)-\rho_{s}(\bxp)\right]
\label{eq:posspace2}
\eea
To see that a Yukawa mean field has been produced, recall Eq.~(\ref{eq:particlecontent}) and take a vacuum expectation value
of this running similarity field. This gives
\be
\langle 0|\phi_s(\bx)|0\rangle=-g\intxp \frac{e^{-m|\bx-\bxp|}}{4\pi|\bx-\bxp|}\left[\rho_{s_0}(\bxp)-\rho_{s}(\bxp)\right]
\ee
which in the high-energy limit ($s\rightarrow s_0$) vanishes and in the low-energy limit becomes
\be
\lim_{s\rightarrow\infty}\langle 0|\phi_s(\bx)|0\rangle=-g\intxp \frac{e^{-m|\bx-\bxp|}}{4\pi|\bx-\bxp|}\rho_{s_0}(\bxp)
\ee
the Yukawa mean field of the similarity dressed source with position space representation given by Eq.~(\ref{eq:rhosx}). 
Among other things note that this mean field is negative (for $g>0$). 

\section{RGPEP}
\label{sec:rgpep}

The seminal and recent papers of RGPEP are given by \cite{RGPEP} and \cite{RGPEP11} respectively. 
Using the machinery of the SRG, and then extending it, the RGPEP method's goal 
is to connect low-energy ``constituent'' particles with their high-energy ``current'' origins. 
In this section we show that a starting equation of the 
RGPEP approach connecting effective Hamiltonians at 
different scales holds exactly in fixed-source neutral scalar theory. 
The utility of this simple model is that this can be done analytically here 
thus giving more confidence in the starting principles. 

To connect to RGPEP, introduce a scalar field that runs in the opposite direction to the one introduced in 
the previous section. Opposite, so $U^\dag(s,s_0)$ acts on the left now instead of the right. The running RGPEP field is 
defined by
\bea
\phi(\bp,s)\equiv
U^\dag(s,s_0)\phi(\bp,s_0)U(s,s_0)&=& \left\{\phi(\bp,s_0)U^\dag(s,s_0)-\left[\phi(\bp,s_0),U^\dag(s,s_0)\right]\right\}U(s,s_0)\nn\\
&=&\phi(\bp,s_0)+g\int_{s_0}^s ds^\pr \rho_{s^\pr}(\bp)\nn\\
&=&\phi(\bp,s_0)+\frac{g}{E_\bp^2}\left[\rho_{s_0}(\bp)-\rho_{s}(\bp)\right]
\label{eq:phisRGPEP}
\eea
which has the opposite sign to Eq.~(\ref{eq:42}), 
but all the other algebra followed as in the previous section. We are not changing the model of this paper in 
this section; we are just seeing how it looks from RGPEP's point of view. 
Note carefully that when the size scale $s$ is used in an argument as in $\phi(\bp,s)$ this 
implies the opposite evolution of RGPEP as in Eq.~(\ref{eq:phisRGPEP}) whereas a subscript as in $\phi_s(\bp)$ 
implies the forward evolution as in Eq.~(\ref{eq:42}). This ``argument $\phi(s)$'' notation is used to connect with 
the notation of \cite{RGPEP11} including its later higher spacetime dimension interpretation (although we do not discuss this further here). 

The running RGPEP conjugate momentum field must also be defined, but since the similarity operator of the model of this paper depends 
only on the conjugate momentum field itself (and not the scalar field), $U(s,s_0)$ and 
$\pi(\bp,s_0)$ commute and we have simply
\be
\pi(\bp,s)\equiv U^\dag(s,s_0)\pi(\bp,s_0)U(s,s_0)=\pi(\bp,s_0)
\label{eq:pisRGPEP}
\ee

In order to show that the RGPEP equation connecting effective Hamiltonians at different scales holds exactly, 
first we need to write the main result of the previous two sections in a new functional notation. 
The previous two sections main result is summarized by 
\bea
H_s[\phi(s_0),\pi(s_0)]&\equiv& U(s,s_0)H_{s_0}[\phi(s_0),\pi(s_0)]U^\dag(s,s_0)\nn\\
&=&\onehalf \intp \left[ \pi^\dag(\bp,s_0)\pi(\bp,s_0) + E_\bp^2 \phi^\dag(\bp,s_0)\phi(\bp,s_0) \right]\nn\\
&&~~~~~~~~~~+ \frac{\Sigma_s}{\sqrt{s}}+g\intp\rho_s(\bp)\phi(\bp,s_0)
\label{eq:theham2b}
\eea
where $\Sigma_s$ and $\rho_s(\bp)$ are the same functions as discussed in the previous two sections. 
The final effective Hamiltonian in RGPEP is defined by taking the just written $H_s[\phi(s_0),\pi(s_0)]$ and 
replacing $\phi(s_0)$ with $\phi(s)$ and $\pi(s_0)$ with $\pi(s)$, 
the oppositely evolving fields of Eqs.~(\ref{eq:phisRGPEP}) and (\ref{eq:pisRGPEP}) respectively. 
In other words, the RGPEP effective Hamiltonian is defined by
\be
H_s[\phi(s),\pi(s)] = H_s[\phi(s_0),\pi(s_0)]|_{[\phi(s_0)\rightarrow\phi(s),\,\pi(s_0)\rightarrow\pi(s)]}
\label{eq:recipe}
\ee
So, what is $H_s[\phi(s),\pi(s)]$ equal to at a different scale, say $s_0$? We show this next.

Following the recipe of Eq.~(\ref{eq:recipe}), the RGPEP effective Hamiltonian becomes
\bea
H_s[\phi(s),\pi(s)]&=&\onehalf \intp \left[ \pi^\dag(\bp,s)\pi(\bp,s) + E_\bp^2 \phi^\dag(\bp,s)\phi(\bp,s) \right]\nn\\
&&~~~~~~~~~~+ \frac{\Sigma_s}{\sqrt{s}}+g\intp\rho_s(\bp)\phi(\bp,s)
\label{eq:theham3b}
\eea
which upon substituting in Eqs.~(\ref{eq:phisRGPEP}) and (\ref{eq:pisRGPEP}) becomes
\bea
&&\!\!\!\!\!\!\!\!\!\!\!\!\!\!\!
H_s[\phi(s),\pi(s)]\nn\\&=&\onehalf \intp \pi^\dag(\bp,s_0)\pi(\bp,s_0) \nn\\
&+& \onehalf \intp E_\bp^2 \left[\phi^\dag(\bp,s_0)+\frac{g}{E_\bp^2}\left\{\rho_{s_0}(\bp)-\rho_{s}(\bp)\right\}\right]
\left[\phi(\bp,s_0)+\frac{g}{E_\bp^2}\left\{\rho_{s_0}(\bp)-\rho_{s}(\bp)\right\}\right]\nn\\
&+& \frac{\Sigma_s}{\sqrt{s}}+g\intp\rho_s(\bp)\left[\phi(\bp,s_0)+\frac{g}{E_\bp^2}\left\{\rho_{s_0}(\bp)-\rho_{s}(\bp)\right\}\right]
\label{eq:theham4b}
\eea
Multiplying this out and also inserting the result for $\Sigma_s$ from Eq.~(\ref{eq:sigmasSfirst1}), 
with real $\rho_s(\bp)$, after the beautiful dust settles, gives
\bea
H_s[\phi(s),\pi(s)]&=&\onehalf \intp \left[ \pi^\dag(\bp,s_0)\pi(\bp,s_0) + E_\bp^2 \phi^\dag(\bp,s_0)\phi(\bp,s_0) \right]\nn\\
&&~~~~~~~~~~+ \frac{\Sigma_{s_0}}{\sqrt{s_0}}+g\intp\rho_{s_0}(\bp)\phi(\bp,s_0)
\label{eq:theham5b}
\eea
with everything at the initial scale $s_0$. We recognize the right-hand side of this last equation as the  
initial Hamiltonian $H_{s_0}[\phi(s_0),\pi(s_0)]$. Thus we have proven the following 
\be
H_s[\phi(s),\pi(s)] = H_{s_0}[\phi(s_0),\pi(s_0)]
\ee
as was to be demonstrated. This is the aforementioned RGPEP equation connecting effective Hamiltonians at different scales. 
Recall that the running RGPEP field is given by Eq.~(\ref{eq:phisRGPEP}).  We now see why the running RGPEP field was in the 
opposite direction: it is a passive renormalization group transformation. The Hamiltonian operator stays fixed in 
the space while the Hamiltonian operator basis fields (and conjugate momentum fields) evolve oppositely. Note that the coordinates are the 
same in the usual active (flow equations and similarity operator) and new passive (RGPEP) approaches, where 
``coordinates'' here means the couplings or coefficients of the Hamiltonian operator, 
$\Sigma_s$ and $\rho_s(\bp)$ in the model of this paper. A simple analogy is an ordinary three-vector rotating in space 
which can be done actively or passively by rotating the vector and keeping the basis vectors fixed or fixing the vector 
and rotating the basis vectors oppositely, respectively; in both cases the final coordinates of this vector are equivalent.

This running RGPEP field satisfies a simple differential equation that we close this section with. Continuing 
the discussion of Eq.~(\ref{eq:phisRGPEP}), we see that the following equation is satisfied in the general similarity model of this paper 
\be
\frac{\partial \phi(\bp,s)}{\partial s} = g\,\rho_s(\bp) 
\ee
So the effective particle field (for $g>0$) is seen to increase with size scale $s$.\footnote{Strictly speaking with 
our conventions a size scale is given by $\sqrt{s}$ as discussed after Eq.~(\ref{eq:posspace1}).} In other words, at low energies the 
effective particles become more pronounced---this is the virtual meson cloud about the source which is distributed 
according to the Yukawa potential. That is one way to look at it.

\section{Summary and Discussion}
Three approaches to the SRG have been presented in this paper: flow equations, similarity operator, and RGPEP. 
Their formal equivalence has been known since the inception of the ideas \cite{G-W, Wegner, RGPEP}. 
Here we analytically show their exact equivalence in fixed-source neutral scalar theory 
with Wegner's canonical generator in the similarity flow equation. This required 
exact cancellation between the free (which became interacting through the similarity operator), 
interacting (original fixed source), and second order background energy (a derived functional of the source) terms. 
In the process, the Yukawa mean field about the source arose 
as in the classic approaches \cite{H-T,Wentzel}, but the regulating static source 
that is required to properly define the theory did not have to be put in by hand but rather followed 
from the similarity flow itself. In addition, a particle picture arose in the last section showing an 
enhancement of the passively evolving RGPEP field at low energies. The methods herein were presented in 
their general forms so as to be extendable to more complicated models.

\begin{acknowledgments}
The authors would like to thank Stan G{\l}azek for discussions. This work was supported in part by the 
National Science Foundation under Grant \mbox{No.~PHY--0653312}.
\end{acknowledgments}

\end{document}